\documentclass[prb,twocolumn,showpacs,superscriptaddress]{revtex4-1}
\usepackage{graphicx}
\usepackage{xcolor}
\usepackage{amsmath}
\usepackage{mathtools}
\usepackage{amssymb}
\usepackage{epsfig}
\usepackage{bm}
\usepackage{color}
\usepackage{tabularx}
\usepackage[normalem]{ulem}
\usepackage[colorlinks,linkcolor=blue,anchorcolor=blue,citecolor=blue,urlcolor=blue]{hyperref}
\usepackage{amsthm}
\usepackage{bbm}
\newcommand{\<}{\langle}
\newcommand{\Fig}[1]{Fig.\ref{#1}}
\newcommand{\Ref}[1]{Ref.\onlinecite{#1}}
\newcommand{\Tr}{{\rm Tr}}
\renewcommand{\>}{\rangle}

\begin{document}

\title{Sub-system fidelity for ground states in one dimensional interacting systems}

\author{Jin-Guo Liu}
\affiliation{National Laboratory of Solid State Microstructures $\&$ School of Physics, Nanjing
University, Nanjing, 210093, China}

\author{Zhao-Long Gu}
\affiliation{National Laboratory of Solid State Microstructures $\&$ School of Physics, Nanjing University, Nanjing, 210093, China}

\author{Jian-Xin Li}
\affiliation{National Laboratory of Solid State Microstructures $\&$ School of Physics, Nanjing University, Nanjing, 210093, China}
\affiliation{Collaborative Innovation Center of Advanced Microstructures, Nanjing University, Nanjing 210093, China}

\author{Qiang-Hua Wang}
\email{qhwang@nju.edu.cn}
\affiliation{National Laboratory of Solid State Microstructures $\&$ School of Physics, Nanjing
University, Nanjing, 210093, China}
\affiliation{Collaborative Innovation Center of Advanced Microstructures, Nanjing University, Nanjing 210093, China}


\begin{abstract}
    We propose to utilize the sub-system fidelity (SSF), defined by comparing a pair of reduced density matrices derived from the degenerate ground states, to identify and/or characterize symmetry protected topological (SPT) states in one-dimensional interacting many-body systems. The SSF tells whether two states are locally indistinguishable (LI) by measurements within a given sub-system. Starting from two polar states (states that could be distinguished on either edge), the other combinations of these states can be mapped onto a Bloch sphere. We prove that a pair of orthogonal states on the equator of the Bloch sphere are LI, independently of whether they are SPT states or cat states (symmetry-preserving states by linear combinations of states that break discrete symmetries). Armed with this theorem, we provide a scheme to construct zero-energy exitations that swap the LI states. We show that the zero mode can be located anywhere for cat states, but is localized near the edge for SPT states. We also show that the SPT states are LI in a finite fraction of the bulk (excluding the two edges), whereas the symmetry-breaking states are distinguishable. This can be used to pinpoint the transition from SPT states to the symmetry-breaking states. 
\end{abstract}

%

\maketitle

\section{Introduction}
\par Identifying the states of matter is one of the central issues of condensed matter physics. Traditional states of matter are characterized by local order parameters in the framework of Landau theory of symmetry breaking.\cite{Landau1958}. However, topological states of matter have no characteristic local order parameters but are topologically nontrivial\cite{Wen2004}. Such states are said to support topological orders. In terms of ground state entanglement, they fall into two categories. The first is the intrinsic topological order\cite{ChenGuWen2010} which has long-range entanglement. It can be characterized by the ground state degeneracy on closed manifolds and the fractional statistics between quasiparticle excitations\cite{ArovasSchriefferWilczek1984,WenNiu1990,LevinWen2003}. The second is the symmetry protected topological (SPT) order\cite{GuWen2009} where the entanglement is only short-ranged. It is a generalization of the topological insulators (TI) and topological superconductors
(TSC)\cite{Hasan2010,Qi2011}, and can be characterized by zero edge modes on open boundaries\cite{Hatsugai1993,RyuHatsugai2002,QiWuZhang2006}.

\par Recently, the diagnosis of SPT states in low dimensions has attracted much attention\cite{}. One strategy is to use the topological quantity constructed from the bulk states. For example, for non-interacting fermionic systems, topological indices such as the Chern number\cite{TKNN1982} and $Z_2$ index\cite{KaneMele2005} can be constructed from bulk energy bands and a complete classification has been developed\cite{SchnyderRyuFurusakiLudwig2008}. However, it was found later that the classification based on non-interacting topological indices may become invalid in the presence of interactions\cite{FidkowskiKitaev2010,FidkowskiKitaev2011}. In such cases the degeneracy in the entanglement spectrum was proposed as an indicator of the non-trivial topology in the bulk wave functions\cite{Pollmann2010,Turner2011}. Meanwhile, by applying matrix-product representation of the wave function, one dimensional SPT phases were shown to be classified by projective representations of
symmetries\cite{Chen2011a,Chen2011b,Pollmann2012}. A more recent scheme is the so-called ``strange correlator''\cite{YouBiRasmussenSlagleXu2014,WuHeYouXuMengLu2015}, which saturates to a constant or decays algebraically if the detected states host non-trivial short-range entanglement. Another way is to detect the nontrivial zero edge modes on open boundaries directly, utilizing the bulk-edge correspondence. In fact, most practical experimental setups\cite{LawLeeNg2009} follow this line since the edge modes are directly measurable. From a theoretical point of view, the implement of this idea is rather obvious for non-interacting fermionic systems because the exact wave function of every single-particle mode can be obtained by solving the single-particle Schrodinger equation. However, in a general interacting system where the single-particle picture does not apply, the verification of the existence of non-trivial zero edge modes is not so straightforward. One possible recipe is to count the degeneracy of the ground states due to the open boundaries, the so-called edge degeneracy. A recent work\cite{WangXuWangWu2015} showed how to detect this edge degeneracy through entanglement entropy. However, a comprehensive investigation is lacking for the identification and construction of the zero edge modes for interacting systems. A related interesting issue is how SPT states are different to the so-called cat states, namely, the symmetry-preserving states by linear combination of symmetry-breaking states. 

\par Here we propose to utilize the sub-system fidelity (SSF), defined by comparing a pair of reduced density matrices derived from the degenerate ground states, to identify and/or characterize SPT states in one-dimensional interacting many-body systems. The SSF tells whether two states are locally indistinguishable (LI) in a given sub-system, a concept developed in quantum information theory.\cite{Fuchs1996,Kells2015}
Starting from two polar states (states that can be completely distinguished on either of the two edges), all the combinations of these states can be mapped onto a Bloch sphere. We prove under generic conditions that a pair of orthogonal states on the equator of the Bloch sphere are LI, independently of whether they are SPT states or cat states. Armed with this theorem, we construct zero-energy exitations in terms of the eigenstates of the reduced density matrix. We show that the zero mode can be located anywhere for cat states, but is localized near the edge for SPT states.  We also show that the SPT states are LI in a finite fraction of the bulk (excluding the two edges), whereas the symmetry-breaking states are distinguishable. This can be used to pinpoint the transition from SPT states to the symmetry-breaking states.     
   
\par The rest of the paper is organized as follows. In Sec.\ref{sec_fidelity}, we introduce the basic idea of SSF. Sec.\ref{sec_examples} shows how it can be implemented in concrete models to differentiate SPT states from cat states. Sec.\ref{sec_summary} provides a summary and perspective remarks. 

\section{Sub-system Fidelity and local indistinguishability}\label{sec_fidelity}
\subsection{Definition}
\par The fidelity $F$ between two ensembles $\text{X}_1$ and $\text{X}_2$ described by the density matrices $\rho_1$ and $\rho_2$, respectively, is defined as,\cite{Fuchs1996}
\begin{equation}
F=\text{Tr}\sqrt{\sqrt{\rho_1}\rho_2\sqrt{\rho_1}}.
\end{equation}
When $\text{X}_1$ and $\text{X}_2$ are pure ensembles, $F$ reduces to the norm of the inner product of the two corresponding quantum states. If $F=1$, the expectation values of any observable with respect to $\rho_1$ and $\rho_2$ are identical. Thus it is the natural generalization of the inner product between two quantum states and measures the overlap of two ensembles: the two ensembles are ``orthogonal'' or distinguishable if $F=0$, and ``identical'' or indistinguishable if $F=1$. 


\par As usual, the reduced density matrix for a pure state $|\psi\rangle$ is obtained by partitioning the system into two parts, say the sub-systems A and B, and then tracing out the degrees of freedom of the environment part B: $\rho_\text{A}=\rm Tr_{\text{B}}|\psi\rangle\langle\psi|$. We obtain two reduced density matrices for two degenerate states, respectively. The resulting fidelity, which we call SSF, depends on the sub-system holding the reduced density matrix. It tells to what extent the two states are indistinguishable by measurements within the given sub-system. 

\par Consider a one-dimensional open chain $\Omega$. We can define, for two given states $|\psi\rangle$ and $|\psi'\>$ on $\Omega$, various SSF's according to the choice of the sub-system. For example, we define $F^{L/R}(l)$ as the fidelity in the left/right part (of the chain) of length $l$. We also define $F^C(l)$ as the fidelity in the central part (symmetric to the two ends) with length $l$. Finally, we define $F^{\bar{C}}(l)$ as the fidelity in the coset $\bar{C}$ of $C$, but with the length of $\bar{C}$ being $l$. In other words, for $\Omega = L\cup C\cup R$, we have $\bar{C}=L\cup R$, with both $L$ and $R$ of length $l/2$. For completeness, we define $F(0)=1$ since any two states are indistinguishable if the sub-system is a null set.

In one-dimension, two states are LI if $F^L(l)\simeq 1$ and $F^R(l)\simeq 1$ for all $l\leq (\mathcal{N}+\xi)/2$, where $\mathcal{N}$ is the total length of the chain and $\xi$ is the characteristic length of local measurements.  On the other hand, $F^C(l<\eta)\simeq 1$ determines the characteristic LI size $\eta$ in the bulk part of the states $|\psi\rangle$ and $|\psi'\rangle$. Such SSF's can be easily implemented in practical calculations.

\subsection{Polar states and locally indistinguishable states}\label{bloch sphere}
\begin{figure}
\includegraphics[width=8.5cm]{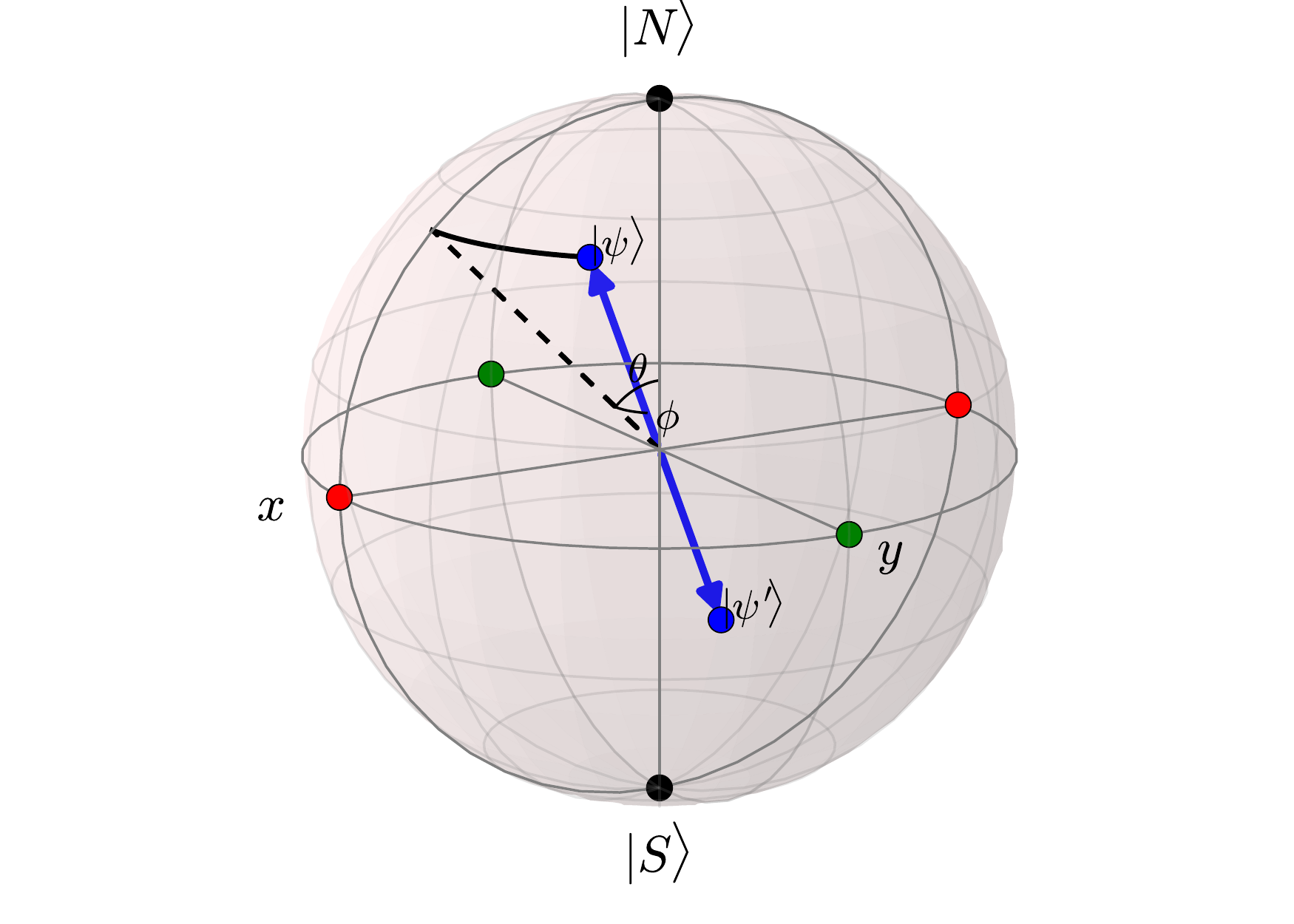}
\caption{Bloch sphere spanned by linear combinations of the edge-distinguishasble states $|N\>$ and $|S\>$ (black dots). The ends of the diameter on the equator, such as the green dots or the red dots, are a pair of locally indistinguishable states.}\label{bloch}
\end{figure}
\par Consider two orthogonal quantum states $|N\rangle$ and $|S\rangle$ which can be distinguished from either edge of the chain, {\em i.e.}, $F^{L,R}_{|N\rangle,|S\rangle}(l>\lambda) = 0$, where $\lambda$ is the characteristic length of the edge. These states can be linearly recombined to form a pair of orthogonal states as
\begin{align}\label{states}
\begin{split}
&|\psi\rangle=\cos\frac{\theta}{2}|N\rangle+e^{i\phi}\sin\frac{\theta}{2}|S\rangle\\
&|\psi^\bot\rangle=\sin\frac{\theta}{2}|N\rangle-e^{i\phi}\cos\frac{\theta}{2}|S\rangle,
\end{split}
\end{align}
where $\theta$ and $\phi$ are Euler angles. They can be mapped onto the so-called Bloch sphere, as illustrated in Fig.\ref{bloch}, with $|N\rangle$ and $|S\rangle$ the polar states at the north and south poles, respectively. Notice that $|\psi\rangle$ and $|\psi^\bot\rangle$ are mapped to the two ends of a diameter of the sphere.

\par We now state and prove a theorem: {\em Any two orthogonal states on the equator of the Bloch sphere are locally indistinguishable, i.e. for $l\in(\lambda,\mathcal{N}-\lambda)$ and $\theta=\pi/2$,  $F_{\psi,\psi^\bot}^{L,R}(l)= 1$.} 
First we perform a Schmidit decomposition $|N\>=\sum_n w_n|n_L\>|n_R\>$ and $|S\>=\sum_s v_s|s_L\>|s_R\>$, where the sizes of the sub-systems $L$ and $R$ are $l$ and $\mathcal{N}-l$, respectively.
By the condition $F^{L}_{N,S}(l)= 0$ and $F^R_{N,S}(\mathcal{N}-l)=0$,
the bases $\{|n_L\>\}$ and $|s_L\>\}$ are orthogonal, and so are $\{|n_R\>\}$ and $\{|s_R\>\}$. In this case, the reduced density matrix for a state from any linear combination of $|N\>$ and $|S\>$ is block-diagonal, for sub-system $\alpha=L/R$,
\begin{align}\label{rhos}
    \begin{split}
    \rho^{\alpha}_\psi=\left(
    \begin{matrix}
        \rho^{\alpha}_N\cos^2\frac{\theta}{2}& \mathbf{0}\\
        \mathbf{0}&\rho^{\alpha}_S\sin^2\frac{\theta}{2}
    \end{matrix}\right),\\
    \rho^{\alpha}_{\psi^\bot}=\left(
    \begin{matrix}
    \rho^{\alpha}_N\sin^2\frac{\theta}{2}& \mathbf{0}\\
        \mathbf{0}&\rho^{\alpha}_S\cos^2\frac{\theta}{2}
    \end{matrix}\right).
    \end{split}
\end{align}
Consequently, it is straightforward to get
\begin{equation}\label{fidelity}
  F_{\psi,\psi^\bot}^{L,R}= \sin\theta,
\end{equation}
where we used the fact that $\Tr\rho^\alpha_{N/S}\equiv 1$. 
Since $\sin\theta=1$ on the equator, we have proven the theorem. 

\par Clearly, the pair of states on the equator is LI in a sub-system up to a length $\mathcal{N}-2\lambda$. 
The theorem implies that we can construct infinitely many pairs of locally indistinguishable states, depending on the phase factor $e^{i\phi}$. On the other hand, since $F_{\psi,\psi^\bot}^{L,R}(l)= 0$ for $\theta=0,\pi$, we see that the polar states can be distinguished by measuring either of two ends. Moreover, we made no assumption on the details of the edges in the polar states. Therefore the theorem holds for both SPT states and cat states. 

A remark is in order. For a finite chain, $F^{L/R}_{|N\rangle,|S\rangle}(l>\lambda)$ can not be exactly zero in general. However, the theorem still holds to a good approximation, as we will show by examples. 

\subsection{Construction of zero modes}\label{sec_zeromode}

We proceed to show how to construct operators that relate two orthogonal states on the equator of the Bloch sphere. For degenerate ground states, they are nothing but zero-energy excitations, or zero modes.
To construct the zero mode at the left edge, we consider a sub-system $L$ of length $l=\lambda$ and its coset $R$ of length $\mathcal{N}-\lambda$. We take advantage of the LI states discussed in the previous subsection,
\begin{align}
    |\phi^{\pm}\>=\sum\limits_n w_n|n_L\>|n_R\>\pm e^{i\phi}\sum\limits_s v_s|s_L\>|s_R\>.
\end{align}
We can then construct an operator on the left edge
\begin{align} \hat{O}_L=\sum\limits_n|n_L\>\<n_L|-\sum\limits_s|s_L\>\<s_L|. \end{align}
By orthogonality between $\{|n_L\>\}$ and $\{|s_L\>\}$, we find $\hat{O}_L|\phi^\pm\>=|\phi^\mp\>$.
Smilarily we can construct the zero mode at the right edge. When the zero mode operators are coupled to a bias field, they break the ground state degeneracy most efficiently. \cite{Liu2011} 

The question is whether the zero mode has to be located near the edge. Take the case of cat states as an example. Since the corresponding polar states $|N\>$ and $|S\>$ are characterized by different local order parameters, they are distinguishable anywhere. We may use any sub-system $\alpha$ in the interior of the chain and the coset $\bar{\alpha}$ in place of $L$ and $R$, respectively. The SSF's for $\alpha$ and $\bar{\alpha}$ satisfy the distinguishability condition for the polar states, therefore we can repeat the above arguments to find that the zero mode between the cat states can also be constructed in a sub-system $\alpha$ of minimal size $\lambda$. Because the location of $\alpha$ is arbitrary, the zero mode can be located anywhere. However, if no local order parameters could be defined except at the edges in the polar states, as in the SPT case, the above $\alpha-\bar{\alpha}$ argument fails. In fact, the bulk part (excluding the two edges) in this case is identical in the polar states, so that there is nothing to be flipped in the bulk. Consequently the zero modes are well-defined only at the edges. In this sense, the SPT states are fundamentally different to the cat states.

\section{Applications}\label{sec_examples}
\subsection{Spin-1 chain}\label{sec_spin1}

\par In this subsection we apply our strategy to the spin-1 chain described by the Hamiltonian \cite{Pollmann2010} \begin{equation}\label{spin1}
    H=\sum_i J\vec{S}_i\cdot\vec{S}_{i+1}-U_{zz}(S^z_{i})^2.
\end{equation}
For an anti-ferromagnetic coupling $J>0$ and a positive $U_{zz}>0$, this model has two phases\cite{}: when $U_{zz}$ is small, the model is in the Haldane phase, which is a topologically non-trivial phase protected by the time reversal symmetry, or the dihedral group of $\pi$ rotations about two orthogonal axes, or bond centered inversion symmetry; when $U_{zz}$ is large, the model is in an anti-ferromagnetic (AFM) phase which breaks the $Z_2^z$ symmetry that flips the $z$ component of the spin operator.

\begin{figure}
\includegraphics[width=8.5cm]{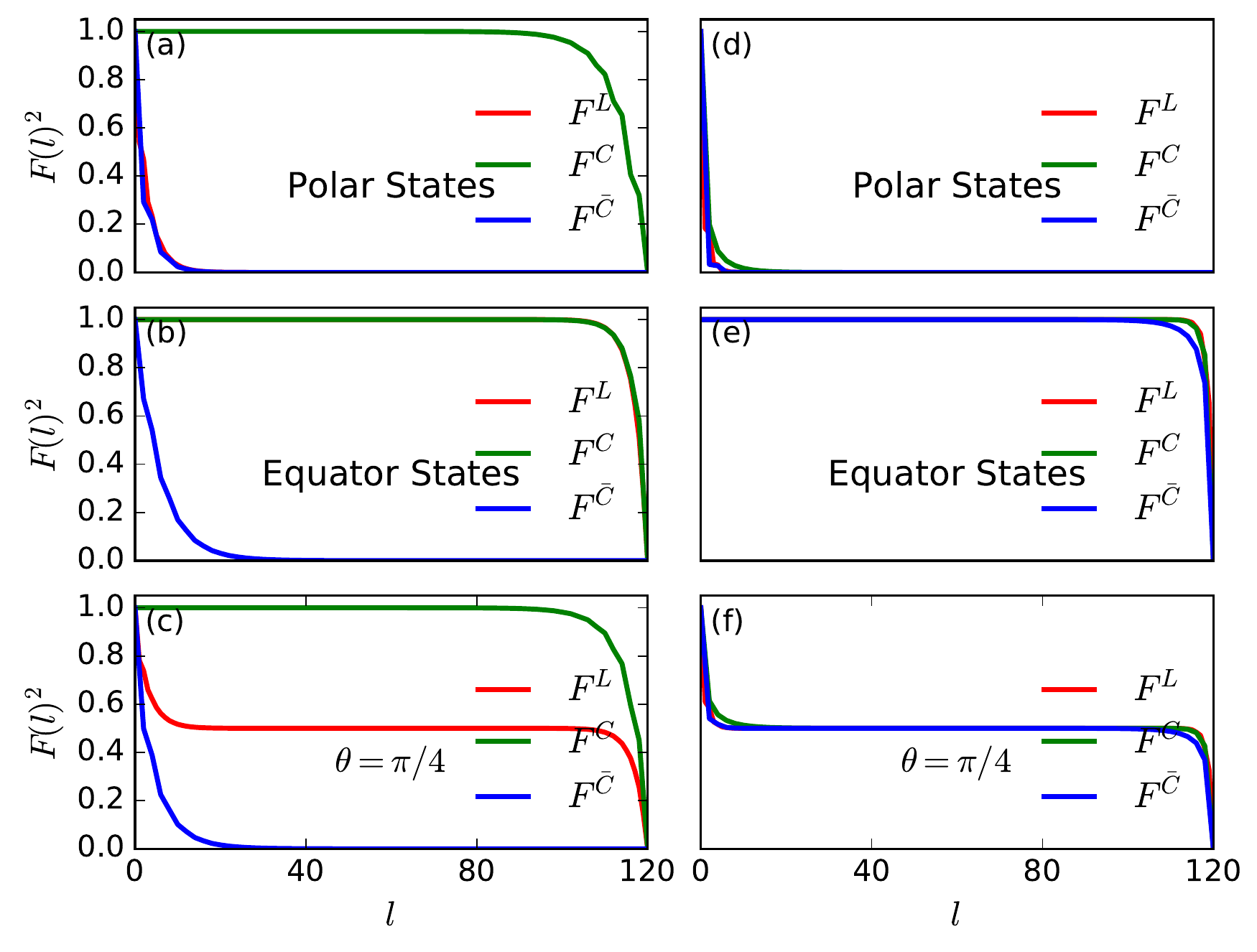}
\caption{SSF versus the sub-system size for a pair of degenerate ground states in the Haldane phase (left panels with $U_{zz}=0$) and AFM phase (right panels with $U_{zz}=1/2$). The pair of ground-states are defined by the ends of a diameter on the Bloch sphere with the Euler angle: $\theta=0$ (polar states) in (a) and (d), $\theta=\pi/2$ (equator states) in (b) and (e), and $\theta=\pi/4$ in (c) and (f), in the respective phases. (See the text for details). Notice that $F^L$ and $F^{\bar{C}}$ in (a) almost coincide, and so do $F^L$ and $F^C$ in (b) and (e).}\label{fidelity_Heisenberg}
\end{figure}
\par We obtain the ground state of this model by the density matrix renormalization group (DMRG) simulations.\cite{Schollwock2005,Schollwock2011} In the calculation, we consider a system with a total number of sites up to $\mathcal{N}=120$ and keep up to $m=80$ states in the DMRG block with more than 5 sweeps to get converged results. The truncation error is of order $10^{-9}$ or smaller.

\par In the open boundary condition, the Haldane phase is 4-fold degenerate, which can be understood by the existence of one isolated spin $1/2$ on each end of the chain. Excluding the identical gapped interior part, these four states can be represented as $|\uparrow\uparrow\rangle$, $|\uparrow\downarrow\rangle$, $|\downarrow\uparrow\rangle$ and $|\downarrow\downarrow\rangle$. Out of these states, $|\uparrow\uparrow\rangle$ lies in the $S_z=1$ sector, $|\downarrow\downarrow\rangle$ in the $S_z=-1$ sector, and $|\uparrow\downarrow\rangle$ and $|\downarrow\uparrow\rangle$ in the $S_z=0$ sector. We consider a submanifold with conserved $S_z=0$. Apparently, $|\uparrow\downarrow\rangle$ and $|\downarrow\uparrow\rangle$ can be distinguished by measurements on either edge because the expectation values of the local operator $S_z$ acting on either end of the chain in these two states are different. Thus these two states define the polar states on the Bloch sphere. Indeed, 
as shown by the red solid line in Fig.\ref{fidelity_Heisenberg}(a), $F^L(l)$ between the polar states $|\uparrow\downarrow\rangle$ and $|\downarrow\uparrow\rangle$ quickly falls to zero as $l$ exceeds a characteristic length. From these polar states, we can define states on the equator as $|X^H_+\rangle=\frac{1}{\sqrt{2}}(|\uparrow\downarrow\rangle+|\uparrow\downarrow\rangle)$ and $|X^H_-\rangle=\frac{1}{\sqrt{2}}(|\uparrow\downarrow\rangle-|\uparrow\downarrow\rangle)$, or $|Y^H_+\rangle=\frac{1}{\sqrt{2}}(|\uparrow\downarrow\rangle+i|\uparrow\downarrow\rangle)$ and $|Y^H_+\rangle=\frac{1}{\sqrt{2}}(|\uparrow\downarrow\rangle-i|\uparrow\downarrow\rangle)$. As is shown in Fig. \ref{fidelity_Heisenberg}(b), $F^L(l)$ is initially unity, and drops to zero only if $l\sim \mathcal{N}$, namely, only if the sub-system includes the two edges. This means the pair of states are locally indistinguishable unless a measurement involving both edges is performed. These pairs of states can be considered as the Bell entangled
states of the edge modes\cite{Leijnse2012}. For orghogonal states with a general $\theta$ on the Bloch sphere, $F^L(l)$ drops from unity initially, saturate at $\sin\theta$ as $l$ is in the bulk, as Eq.\ref{fidelity} requires, and drops to zero when the right edge is approached, as shown in Fig. \ref{fidelity_Heisenberg}(c) for $\theta=\pi/4$. Exactly the same behavior is observed for $F^R(l)$, with the understanding that the sub-system starts from the right edge and expands (as $l$ increases) toward the left edge.

\par In the AFM phase, the ground state is 2-fold degenerate. We use $|+\rangle$ and $|-\rangle$ to represent the two polar states. They are apparently distinguishable at any site. We can form cat states $|X^{AFM}_+\rangle=\frac{1}{\sqrt{2}}(|+\rangle+|-\rangle)$ and $|X^{AFM}_-\rangle=\frac{1}{\sqrt{2}}(|+\rangle-|-\rangle)$, or $|Y^{AFM}_+\rangle=\frac{1}{\sqrt{2}}(|+\rangle+i|-\rangle)$ and $|Y^{AFM}_+\rangle=\frac{1}{\sqrt{2}}(|+\rangle-i|-\rangle)$. As shown in Fig. \ref{fidelity_Heisenberg} (right panels), $F^L(l)$ for the polar (d), equator (e) and $\theta=\pi/4$ (f) states in the AFM phase is similar to the corresponding cases in the Haldane phase (left panels). 

\par However, the SSF $F^C(l)$ (for a central sub-system of size $l$) behaves rather differently, as shown by the green solid lines in Fig.\ref{fidelity_Heisenberg}. In the Haldane phase (left panels), it falls to zero only if $C$ touches the ends of the chain, independently of the Euler angle $\theta$, while in the AFM phase (right panels), only the equator states, or cat states, in (e) show a similar behavior. In fact, this is a common phenomenon for the SPT phases and the symmetry-breaking (SB) phases in one dimension. The SPT phase has a unique non-degenerate ground state in the periodic boundary condition, thus the difference between the degenerate ground states only occurs at the edges in the open boundary condition. However, the SB phase has degenerate ground states even in the periodic boundary condition, and the difference between these SB states (as polar states) is measurable everywhere. But this also implies that there is no difference between bulk and edge in the cat states, which are therefore indistinguishable in $C$, as $F^C$ shows in (e). Taking advantage of the difference for the polar states, we can use the normalized bulk indistinguishable length $\eta/\mathcal{N}$, with $\eta$ extracted from $F^C(l\leq \eta)\sim 1$ or $F^C(l\geq \eta)\sim 0$ for the polar states, as a representative parameter to identify the SPT phases.

The difference also occurs in SSF $F^{\bar{C}}(l)$ for a sub-system $\bar{C}=L\cup R$ out of $\Omega = L\cup C\cup R$, shown as blue solid lines in Fig.\ref{fidelity_Heisenberg}. In the Haldane phase (left panels), $F^{\bar{C}}(l)$ drops to zero quickly, implying that the states can be distinguished immediately by non-local measurements involving both edges. For the polar states, this is obvious since different spin moments are present on the edges. In the SPT states, the edge spins form either a singlet or a triplet, and this is distinguishable if both edges are simultaneously measured. As $\bar{C}$ increases toward the center, no further change can be anticipated since the bulk is identical in both SPT states. The states with $\theta=\pi/4$ lie inbetween the two extremes and thus behave similarly. In the AFM phase (right panels), $F^{\bar{C}}(l)$ in Fig.\ref{fidelity_Heisenberg}(d) also drops quickly for the polar states, since the latters are distinguishable everywhere by local order parameters. In contrast to the SPT states, however, Fig.\ref{fidelity_Heisenberg}(e) (blue line) shows the cat states (equator states) can not be distinguished in $\bar{C}$ unless $l\simeq\mathcal{N}$. This is consistent with the previous argument that there is no difference between edge and bulk in the cat states, hence there is no qualitative difference among $F^{L,C,\bar{C}}$ in Fig.\ref{fidelity_Heisenberg}(e). Finally  Fig.\ref{fidelity_Heisenberg}(f) can be understood as an interpolation between (d) and (e). 

\begin{figure}
\includegraphics[width=8.5cm]{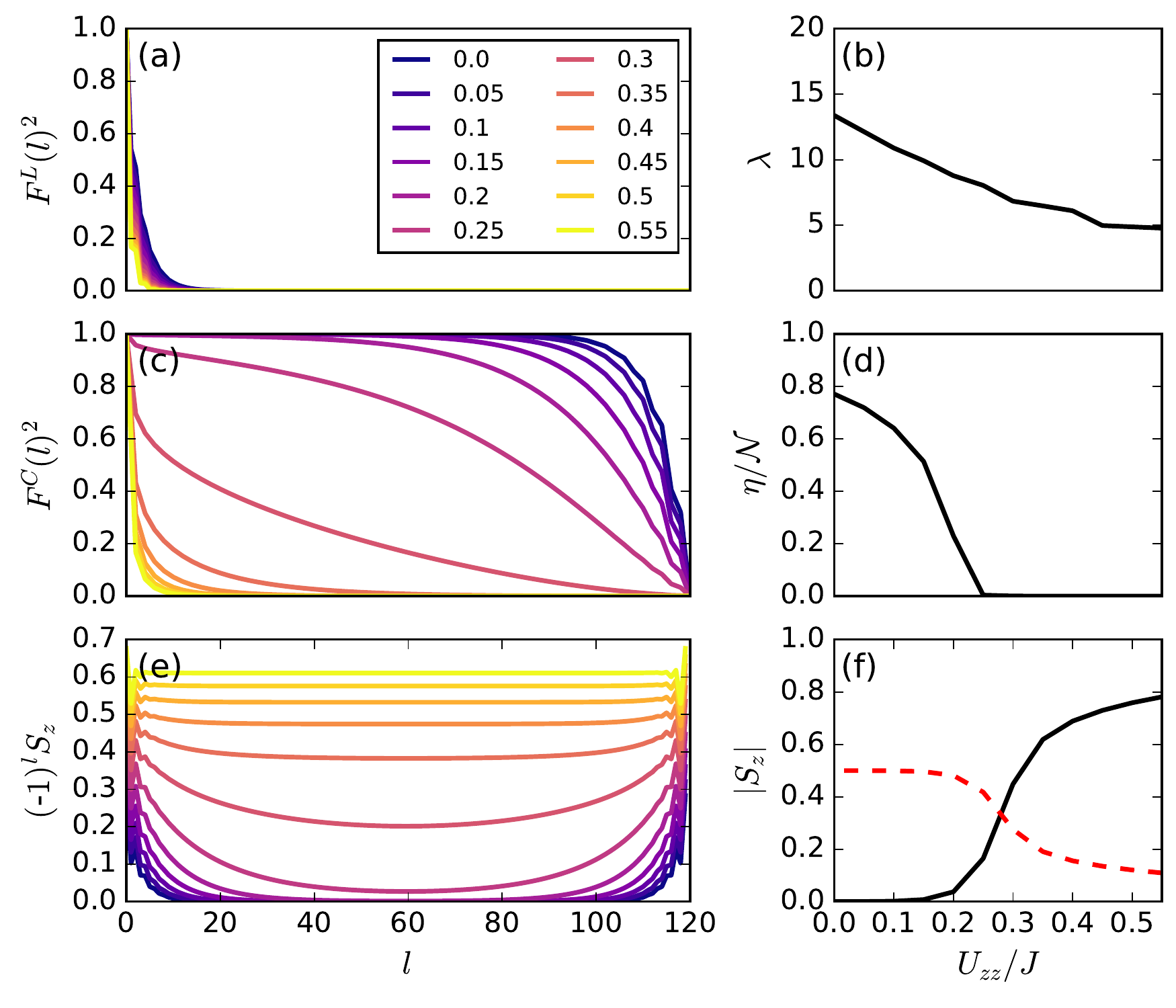}
\caption{Numerical results in the $S_z=0$ sector for a range of $U_{zz}$ indicated in the legend. (a) Plot of $F^{L}(l)$. (b) The penetration depth $\lambda$ of the edge modes determined from (a). (c) The same plot as (a) but for $F^C(l)$. (d) The bulk indistinguishable length $\eta$ determined from (c), plot as $\eta/\mathcal{N}$. (e) The anti-ferromagnetic moment on the chain. (f) The anti-ferromagnetic moment (black solid line) on the central two sites of the chain and the net magnetic moment (red dashed line) in the left half of the chain.}\label{phasetran}
\end{figure}

\par Having characterized the SPT and SB phases separately, we now discuss how they are transformed into each other as the control parameter $U_{zz}$ varies in a physical system. The results are presented in Fig.\ref{phasetran}. From $F^L(l)$ in (a) we extract, by a criterion $|F^L(l\geq \lambda)|^2<10^{-2}$, the penetration depth $\lambda$ of the edge modes, which is plot versus $U_{zz}$ in (b). Similarly, from $F^C(l)$ in (c) we extract, by a criterion $|F^C(l\geq \eta)|^2<10^{-2}$, the bulk indistinguishable length $\eta$, plot as $\eta/\mathcal{N}$ versus $U_{zz}$ in (d). In the thermodynamic limit, a finite $\eta/\mathcal{N}$ implies bulk indistinguishability. With the increase of $U_{zz}$, $\lambda$ decreases slightly but keeps to be nonzero, while $\eta/\mathcal{N}$ is finite for small $U_{zz}$ but drops to zero at $U_{zz}\simeq J/4$. This turns out to be in agreement with the transition point found in \Ref{Pollmann2012}. Therefore $\eta/\mathcal{N}$ is a useful indicator of the SPT. We check how this corresponds to the distribution of the average local moments. As shown in Fig. \ref{phasetran}(e), (in the polar states) the local moments appear only at the edge in the Haldane phase where $U_{zz}<J/4$, while they appear everywhere in the SB AFM phase where $U_{zz}>J/4$.  With increasing $U_{zz}$, we see a gradual penetration of local moments from the edge to the bulk. In Fig. \ref{phasetran}(f), we show the local moments at the two central sites (the black solid line) and the net magnetic moment in the left half of the chain (the red dashed line). 
(The expectation values are identical, up to a minus sign, in each pair of orthogonal and degenerate states.)
Just as expected, the AF moment develops a non-zero value around $U_{zz}\simeq J/4$, and at the same time the net magnetic moment is about $1/4$ (half way between zero and half spin). These results indicate a transition point $U_{zz}\simeq J/4$, in agreement with that determined by $\eta/\mathcal{N}$.

\subsection{Interacting Kitaev chain and XZ spin model}
\par The Kitaev chain\cite{Kitaev2001} describes spinless fermions with $p$-wave pairing in one dimension. The original Kitaev model is free from interaction. Here we incooporate interaction to investigate the many-body effects. The Hamiltonian is
\begin{eqnarray}\label{Kiteav}
  H=&&\sum_i (tc_i^\dagger c_{i+1}+\Delta c_i^\dagger c_{i+1}^\dagger+{\rm h.c.})-\mu \sum_i c^\dagger_ic_i\nonumber\\ &&+V\sum_i (n_i-1/2)(n_{i+1}-1/2).
\end{eqnarray}
Here $t$, $\mu$ and $V$ are hopping integral, chemical potential and nearest-neighbor repulsive interaction, respectively. For simplicity, we limit ourselves to the case $t=\Delta$. When $|\mu|\ll t$ and $V\ll t$, this model is in the topological phase which hosts an unpaired Majorana fermion on each edge. Through Jordan-Wigner transformation, the model can be mapped to an XZ spin model,
\begin{equation}\label{XZ}
  H=\sum_i (J_xS^x_iS^x_{i+1}+J_zS^z_iS^z_{i+1}-hS_i^z),
\end{equation}
with $J_x=4t$, $J_z=V$ and $h=\mu$.
\par The ground states of these two models are determined by the DMRG simulations in a similar setting described above. In the calculation of SSF in the Kitaev model, we always bring the target sub-system to the left of the environment to avoid the fermion sign caused by the environment. (The fermion sign within each sub-system is retained rigorously.)

\begin{figure}
	\includegraphics[width=8.5cm]{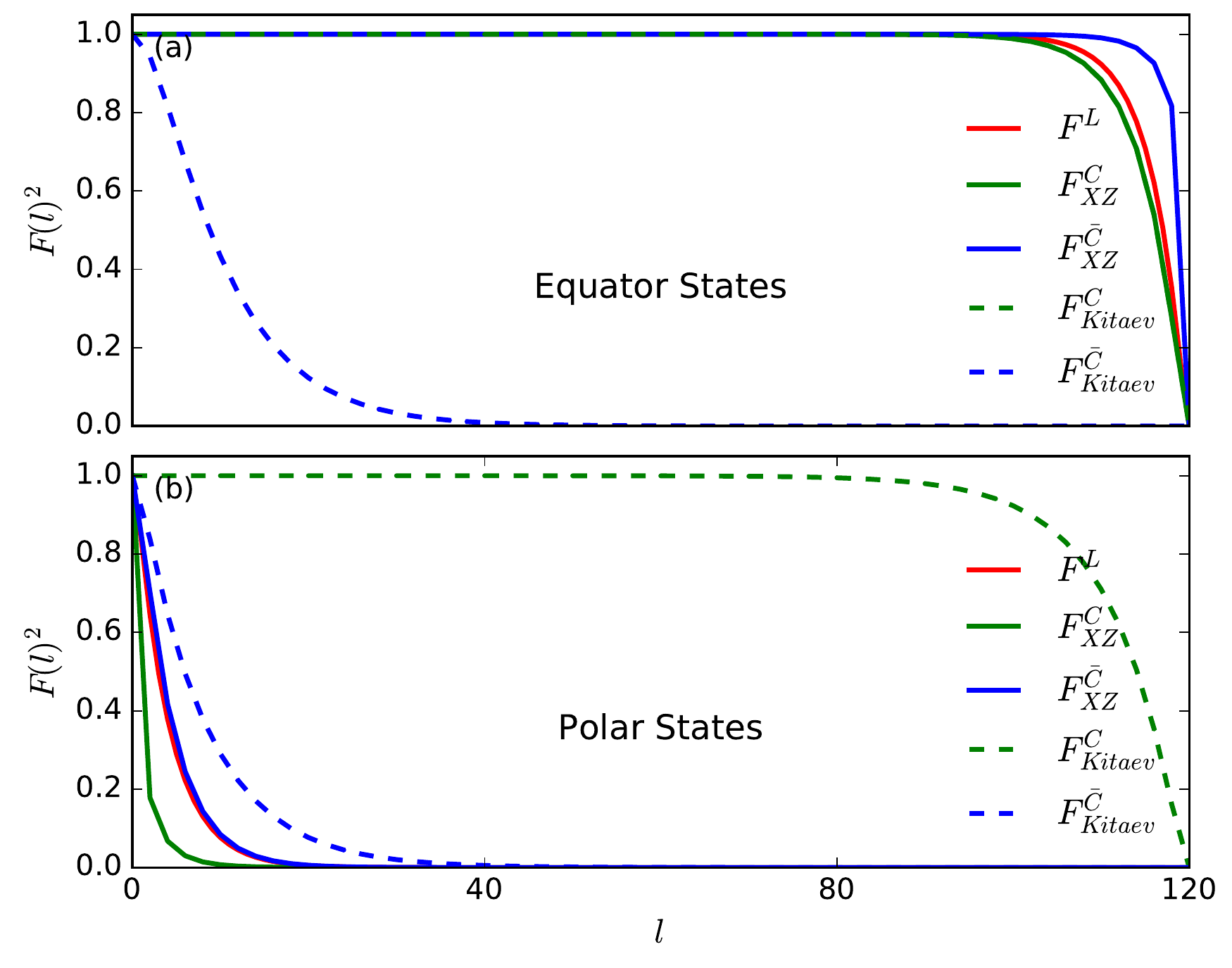}
	\caption{(a) SSF between pairs of symmetry-preserving states in the Kitaev model (dashed lines) and the XZ model (solid lines). (b) The same plot as (a) but for polar states.}\label{fidelity-kitaev}
\end{figure}

\par By correspondence, the topological phase of the Kitaev model is mapped to the $Z_2^z$ SB phase (the $S_x$-ordered phase) of the XZ model. The latter was often argued to also host Majorana zero modes in the literature.\cite{Vishveshwara2011,Bardyn2012,Zvyagin2013} However, since the Jordan-Wigner transformation is  non-local, the physics of these two phases is rather different, even though they are mathematically equivalent.\cite{Greiter2014} As shown in \Fig{fidelity-kitaev}(a), for symmetry preserving states (equator states), these two phases show exactly the same features in $F^{L}(l)$: both of them are LI. However, differences take place when measurements are performed on both edges by tracing out $C$ in favor of $F^{\bar{C}}(l)$. With moderate $l$, the cat states in the XZ model are indistinguishable, while the SPT states in the Kitaev model can certainly be distinguished. In \Fig{fidelity-kitaev}(b) we show SSF's in the polar states. They can be distinguished in all cases, except for the case in $C$ for the Kitaev model. The underlying reason for the above results is the presence/absence of local order parameters in the bulk, as we discussed for the spin-1 Heisenberg model. On the other hand, the zero energy excitation can be located within any continuous segment in the XZ model, as pointed out in Sec.\ref{bloch sphere}. In this sense, we believe it is improper to call the edge operators in the XZ model as genuine edge modes, although they could be mapped to Majorana operators through Jordan-Wigner transformation. In fact, if the fermion sign in the Kitaev model were ignored in the analysis of the (true) ground states, we would get exactly the same SSF's for both models.

\section{Summary}\label{sec_summary}
Utilizing the sub-system fidelity, we proved both analytically and numerically that degenerate symmetry preseving states (equator states) in one-dimensional models can not be distinguished locally, for both SPT states and cat states. Taking advantage of SSF, we show how local zero mode operator swapping the degenerate states can be constructed. The zero mode can be located anywhere in the SB phase, while it can only be located at the edges in the SPT case. Finally we show that the bulk indistinguishable length extractable from SSF can be used to pinpoint the phase transition from the SPT state to the SB state. 

The SSF could be extended to excited states to look for their characteristics. It may also be extended to higher dimensions. We leave these posibilities in future studies.

\acknowledgments{The project was supported by NSFC (under grant Nos.11574134 and 11374138) and the Ministry of Science and Technology of China (under grant No. 2016YFA0300401). LJG thanks D. Wang and S. R. Manmana for helpful discussions.}

\section{Appendix}\label{app_numerical}


For a given state $|\psi\>$, we perform a Schmidt decomposition,
\begin{equation}
|\psi\> = \sum_{n}\alpha_{n} |n_L\>|n_R\>,
\end{equation}
where $L$ is a sub-system and $R$ the environment. We can trace out $R$ in favor of a reduced density matrix in $L$,
\begin{equation} \rho^L = \sum_{n,m} \alpha_n\alpha_m^*|n_L\>\<m_L| = \sum_k \Lambda_k^2 |k\>\<k|.\end{equation}
In the last equality we write the density matrix in the diagonal basis. For two states $|\psi_1\>$ and $|\psi_2\>$, we get two corresponding density matrices $\rho^L_1$ and $\rho^L_2$, and this can be used to calculate the SSF,
\begin{eqnarray} F^L &&= \Tr\sqrt{\sum_{k_1 k_2 k_{1'}}\Lambda_{k_1}\Lambda_{k_2}^2\Lambda_{k_{1'}} |k_1\>\<k_1|k_2\>\<k_2|k_{1'}\>|k_{1'}|}\nonumber\\ &&=\Tr\sqrt{\mathcal{O}\mathcal{O}^\dagger},\end{eqnarray}
where $\mathcal{O}$ is the overlap matrix,
\begin{equation} \mathcal{O}_{k_1,k_2} = \Lambda_{k_1}\Lambda_{k_2}\<k_1|k_2\>.\end{equation}
Numerically we keep the MPS right canonical, namely, the states in $R$ form an orthonomal set.
A final singular-value decomposition $\mathcal{O} = U S V$ enables us to write
\begin{equation} F^L = \Tr~ S.\end{equation}
Notice that $F^L$ depends on $|\psi_1\>$, $|\psi_2\>$ and the size $l$ of $L$. \\

\begin{figure}
	\includegraphics[width=8.5cm]{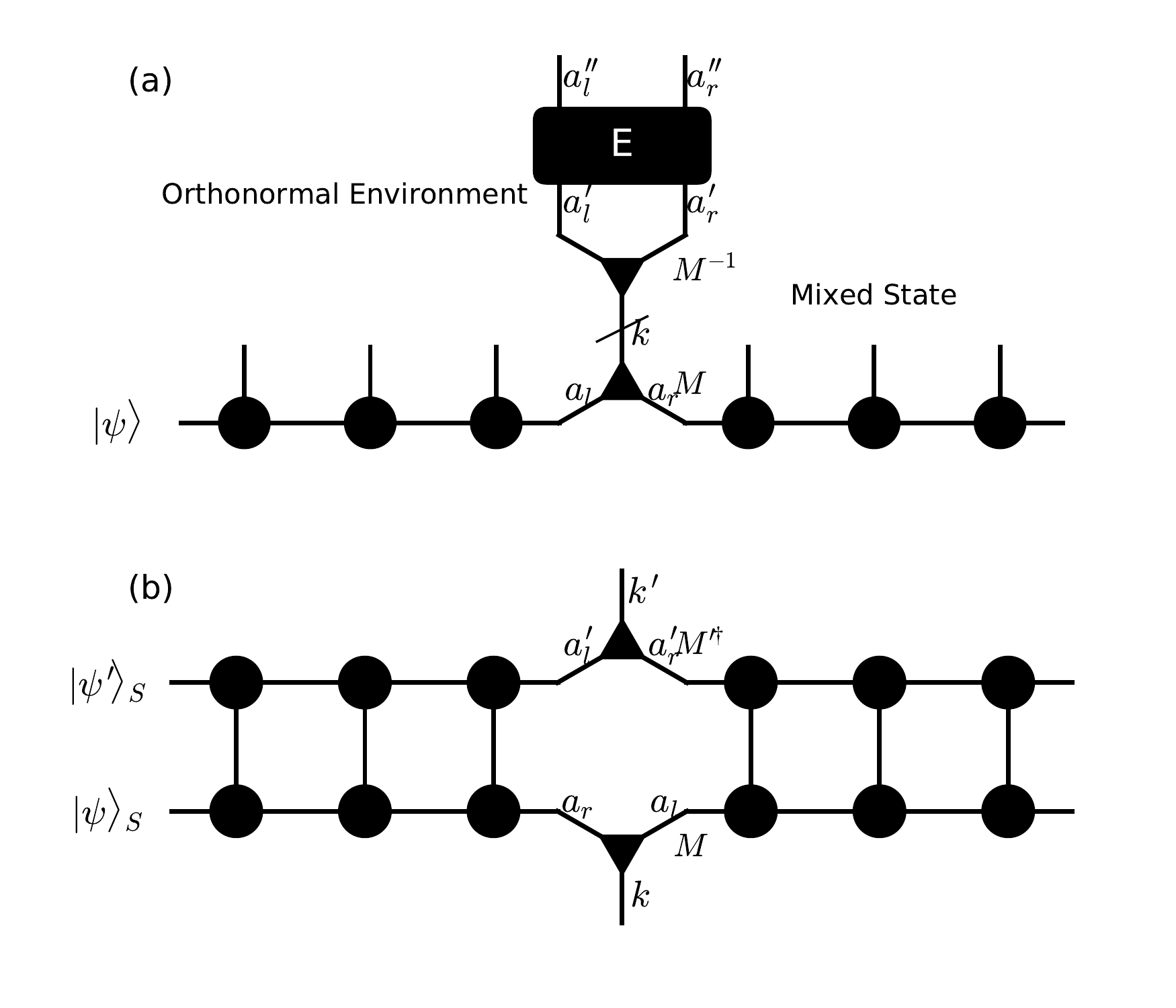}
	\caption{Illustration of (a) how the quantum mixtures are formed, and (b) how fidelity between two states are calculated using matrix product representation. Here, the circles are real sites, the rounded rectangle is the overlap matrix of central block $M$, and the Cholesky decomposition of $E$ is denoted by the  triangles connected by a bond.}\label{mixture}
\end{figure}

However, the evaluation of the overlap matrix becomes tricky for $F^{\bar{C}}$, where $\bar{C}=L\cup R$ out of $\Omega=L\cup C\cup R$. The difficulty is to make the states in the environment block $C$ orthonormal.
The idea is to get a mixed state by tracing out the environment $C$.
In the first stage, we get the outer product for the environment block 
\begin{equation} E_{a_l a_r,a_l'a_r'}=\sum\limits_{\{\sigma_c\}}\Psi_{a_l,a_r}^{\{\sigma_c\}}\Psi^{\{\sigma_c\}}_{a_l',a_r'}\end{equation}
where $\Psi_{a_l,a_r}^{\sigma_c}$ is the matrix representation of the environment, with $a_{l,r}$ the virtual indices and $\{\sigma_c\}$ the physical indices to be contracted.
Then we perform a Cholesky decomposition for $E$ to get 
\begin{equation} E_{a_la_r,a_l'a_r'} = \sum_{k=1}^K M_{a_la_r,k} M^\dagger_{k,a_l'a_r'},\end{equation}
where $K$ is the number of composite indeces $a_l a_r$. Thus we have, in matrix form,  $M^{-1}EM^{-1\dagger}=\mathbbm{1}$ (\Fig{mixture}(a)). This implies that the basis set $\{|k\>\}$ defined by
\begin{equation} \<\{\sigma_c\}|k\> = \sum_{a_l a_r} M^{-1}_{k,a_l a_r}\Psi_{a_l a_r}^{\{\sigma_c\}}, \end{equation}
forms an orthonormal basis set.   
At last, we insert $MM^{-1}$ into the original MPS between environment and system blocks, and trace out the environment reexpressed in the basis $\{|k\>\}$. This is equivalent to getting rid of the upper part above the slash in \Fig{mixture}(a), leaving a tensor coefficient $M_{k,a_la_r}$.
In this way we get the MPS representation of a mixed state in the system block.
Notice that in the Cholesky decomposition, the number $K^*$ of nonzero $M$'s is bounded by $K^*\le \min(d^{n_E},K)$ where $n_E$ the number of sites in the evironment block and $d$ the dimension of the single-site Hilbert space. Rank defficiency (or $K^*<K$) occures if $d^n_E<K$, making $M^{-1}$ ill defined.
However, we find this is harmless since $M^{-1}$ does not appear in the final representation. In practice, we use eigenvalue decomposition instead to get around the possible difficulty in the Cholesky decomposition.
Finally, the overlap matrices $\mathcal{O}_{k,k'}(\psi,\psi')$ can be obtained by contracting the pysical indices between two mixed states, see \Fig{mixture}(b) for illustration.

We remark that the labeling of sites in fermionic systems matters because the parity of electrons in the environment blocks trapassed by the system block may cause a negative sign. To avoid such a fermion sign problem, we always label the system blocks first, and the environment second. Notice that fermion signs are respected rigorously within the system and environment, respectively. 

\bibliography{topo}

\end{document}